\documentclass[prb,twocolumn,showpacs,preprintnumbers,amsmath,amssymb]{revtex4}
\usepackage{graphicx}

\begin{document}

\title{Non-hexagonal-ring defects and structures induced by strain in
graphene and in functionalized graphene}

\author{Joice da Silva-Ara\'ujo} 

\author{A. J. M. Nascimento}

\author{H\'elio Chacham} 

\author{R. W. Nunes}
\email{rwnunes@fisica.ufmg.br}

\affiliation{Departamento de F\'{\i}sica,  Universidade Federal de Minas
Gerais, CP 702, 30123-970, Belo Horizonte, MG, Brazil}

\begin{abstract}

We perform {\textit ab initio} calculations for the strain-induced
formation of non-hexagonal-ring defects in graphene, graphane (planar
CH), and graphenol (planar COH). We find that the simplest of such
topological defects, the Stone-Wales defect, acts as a seed for
strain-induced dissociation and multiplication of topological
defects. Through the application of inhomogeneous deformations to
graphene, graphane and graphenol with initially small concentrations
of pentagonal and heptagonal rings, we obtain several novel stable
structures that possess, at the same time, large concentrations of
non-hexagonal rings (from fourfold to elevenfold) and small formation
energies.

\end{abstract}

\maketitle

\section{Introduction}
The intense interest in the Physics of graphene, over the last few
years, derives in large part from the exceptional electronic
properties of the pristine material.~\cite{rmp,novos07,novos04,kats06}
While pristine samples obtained by exfoliation of graphite have allowed
exploration of a rich variety of phenomena connected with the
Dirac-fermion nature of the electronic excitations in graphene,
current large-scale synthesis protocols have reached the point where
polycrystalline or highly-inhomogeneous samples are being
produced.~\cite{cristina,kotakoski,westenfelder,bagri} In very recent
experiments, graphene layers resulting from reduction of graphene
oxide show conductivities that are 1 or 2 orders of magnitude smaller
than those of pristine
graphene,~\cite{jung,gilje,navarro,lopez,tung,eda,wu} possibly due to
residual dopants and intrinsic defects. Better conductivities can be
obtained when partially oxidized graphene is used as a
precursor.~\cite{goki}

From the perspective of applications, widespread use of graphene will
depend on the control and understanding of the properties of such
defective graphene products. The atomic structure of defects in
reduced graphene oxide (RGO) has been recently investigated through
high resolution transmission electron microscopy
(TEM)~\cite{cristina,goki} and molecular dynamics
simulations.~\cite{bagri} The experimental results show significant
areas with holes, grain boundaries, isolated pentagon-heptagon pairs
and clusters of carbon pentagons, hexagons, heptagons and a few
octagons. Similar defect morphologies can be generated in exfoliated
graphene by electron-beam irradiation,~\cite{kotakoski} and by
irradiation under extreme temperatures.~\cite{westenfelder} The
occurrence of areas with high density of extended topological defects
(ETD) observed in the samples of Ref.~5 was
attributed to the process of oxidation and further reduction of the
graphene oxide, since such large defective areas do not appear in
mechanically exfoliated graphene from the same graphite
source. Moreover, in these samples residual functional groups were
found to concentrate primarily on regions with ETDs, surrounded by
pristine (unfunctionalized) graphene areas.

Given this current interest in defective graphene systems, in the
present work, we employ {\it ab initio} density functional theory
(DFT) calculations to study the morphological evolution of the
following graphene structures: (i) homogeneously-strained graphene
sheets containing a Stone-Wales (SW) defect;~\cite{SW} and (ii)
pristine and functionalized graphene sheets containing initial
distributions of isolated topological defects (TD),~\cite{jrc} after
introduction of holes and highly inhomogeneous bond deformations. We
find that stress-relaxation of a graphene sheet containing SW defects
may lead to the formation of ETDs, with morphological units that are
very similar to those observed in the aforementioned
experiments.~\cite{cristina,kotakoski,westenfelder,bagri} We uncover
the role of the SW defect as a stress-accumulation site that induces
the bond rotation events which generate the ETDs. Furthermore, we find
that healing of small voids coupled with the relaxation of
inhomogeneously strained regions also lead to the formation of ETDs,
revealing a rich variety of morphological patterns and plastic
deformation mechanisms in pure and functionalized graphene
structures. Our results indicate that the marked tendency of graphene
to form TD clusters under strain is aided by functional groups, since
the formation energies of such defective structures are considerable
smaller in graphane (a graphene sheet fully functionalized with
hydrogen atoms) and in graphenol (a graphene sheet fully functionalized
with hydroxyl groups) than in graphene, in agreement with the
experimental evidence in Ref.~5.

Our calculations are performed using Kohn-Sham density functional
theory,~\cite{Kohn} the generalized-gradient approximation
\cite{Kleinman} for the exchange-correlation functional, and
norm-conserving Troullier-Martins pseudopotentials,~\cite{Troullier}
to describe the electron-ion interactions. We use the LCAO method
implemented in the SIESTA code,~\cite{Siesta} with a double-zeta
pseudo-atomic basis set augmented with polarization orbitals, with an
energy cutoff of 0.01 Ry. Full structural optimization of atomic
positions and supercell vectors is performed. For the relaxed
structures the total force on each atom is less than 0.02~eV/\AA\ and
the pressure on the supercell is less than 1~kBar. The supercells
employed are periodic along the graphene plane, being surrounded by a
30~\AA\ vacuum region in the transversal direction, with negligible
interactions between each layer and its periodic images.

\section{Results and discussion}
\subsection{Homogeneous shear deformations of graphene containing Stone-Wales defects}
\subsubsection{First principles results}
\begin{figure}[b]
\includegraphics[width=8.5 cm]{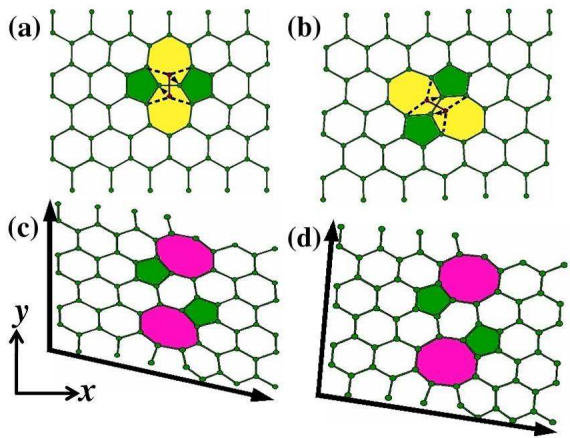}\\
\caption{Morphological transformation of a graphene sheet under
strain.  (a) and (b) show two images of graphene
supercells containing a Stone Wales defect. The bond rotation that
generates the defect is indicated. The two cases differ by the
orientation of the defect with respect to the $x$ and $y$ axes of the
cell. (c) Transformation of the supercell in (a) with relaxation of
internal atomic coordinates, under homogeneous shear strain. (d)
Fully-relaxed stress-free state of the structure in (c), after
relaxation of cell vectors.}
\label{fig1}
\end{figure}

We consider initially a likely scenario, where homogeneous shear
deformations are imposed upon a graphene sheet containing a SW
defect. SW defects are commonly found in graphene and nanotubes, and
can be generated by irradiating the material with an 80 keV electronic
beam, just below the knock-on threshold for $sp^2$ bonded carbon
materials.~\cite{kotakoski,meyer,banhart} The impinging electrons give
out the energy needed for the 90$^\circ$ rotation of a carbon-carbon
(C-C) bond [as indicated in Figs.~\ref{fig1}(a) and (b)] that
transforms four hexagons of the pristine graphene matrix into a SW
defect. Since a SW is an incipient dislocation dipole in graphene
(each pentagon-heptagon pair is one of two dislocations with opposite
Burgers vectors comprising the SW defect), at first sight one expects
shear deformations either to undo the C-C bond rotation and thus lead
to self-annihilation of the two pentagon-heptagon pairs, or to lead to
dissociation of the dislocation pair into its individual components.

In order to investigate this, we consider the two graphene supercells,
each containing a SW defect, shown in Fig.\ref{fig1}(a) and (b). They
differ by the orientation of the SW with respect to the $x$ and $y$
axes of the cell (indicated in the figure). We find that a 30\% shear
strain of the supercell shown in Fig.~\ref{fig1}(a) leads to the
formation of the ETD shown in Figs.~\ref{fig1}(c) and (d), containing
clusters of fivefold and eightfold rings, while no morphological
transformation is observed when the same deformation is imposed on the
supercell shown in Fig.\ref{fig1}(b). We shall point out that
previous works have found pristine graphene to sustain deformations of
up to 25\% without yielding,~\cite{lee,cadelano,min} a result that we
also find in our calculations, for graphene with SW defects. The
morphological pattern observed in Fig.~\ref{fig1}(c) is obtained by
allowing the system to fully relax all the internal atomic
coordinates, while maintaining the imposed homogeneous strain.  No
further changes in morphology are observed when full relaxation of
both atomic coordinates and strain (i.e. the supercell vectors) are
allowed, as shown in Fig.~\ref{fig1}(d).

These results highlight a fundamental aspect of the plastic response
of a graphene sheet to imposed stresses or strains: a tendency for the
formation of extended topological defects as a stress-relaxation
mechanism, spawned by bond rotation events, in the above example with
the formation of clusters of fivefold and eightfold rings. We note
also that the transformation depends on the orientation of the shear
deformation with respect to the preexisting SW defect, that acts as
stress-accumulation seed for the topological transformation.

\subsubsection{Analysis of elastic energy using a Keating model}
\begin{figure}[b]
\includegraphics[width=8.5 cm]{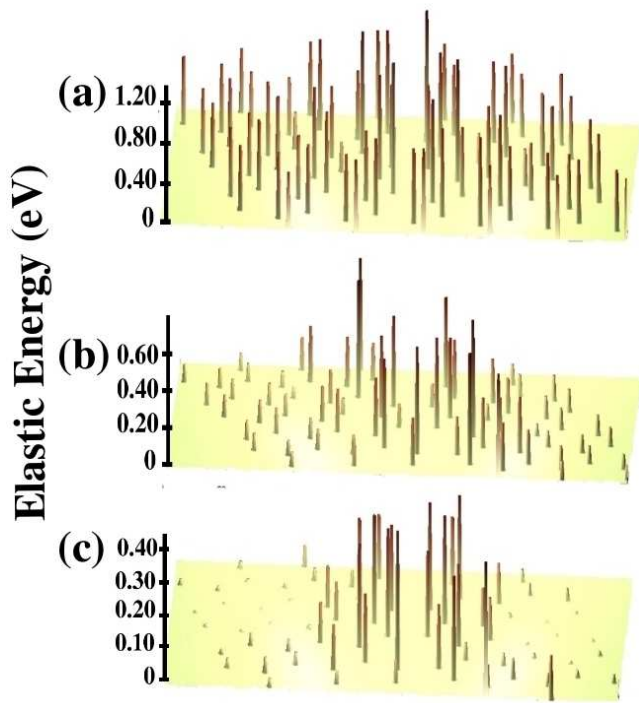}\\
\caption{Keating-energy analysis of stress relaxation by extended
topological defect formation. In the initial structure a homogeneous
shear strain is imposed on a graphene cell containing a Stone-Wales
defect, as shown in Fig.~\ref{fig1}(a). The location of the defect is
at the top, along the center of the cell, in the regions with higher
elastic-energy histogram bars. (a) In the initial structure, the
elastic-strain energy is nearly uniform in both defective and bulk
regions of the cell. (b) Concentration of elastic energy, after
morphological transformation that generates an extended topological
defect, in the process of stress relaxation under applied strain.  (c)
Relaxation of bulk and defect strain in the stress-free metastable
structure. Most of the residual strain is located on the defect.}
\label{fig2}
\end{figure}

The latter point can be illustrated by following the evolution of the
elastic energy of the cell during the deformation, as shown in
Fig.~\ref{fig2}. In the figure, we show a map of the elastic energy of
the supercell at three different stages of the plastic relaxation
process, computed using a Keating model.~\cite{keating}
Figure~\ref{fig2}(a) shows the elastic energy for the initial strained
geometry. Note that, in this configuration, the strain energy is
distributed nearly uniformly over the whole system, with a larger
magnitude at the SW defect. This pattern is preserved up to the point
where a morphological transformation takes place. Figure~\ref{fig2}(b)
shows the strain energy right after the transformation: the important
feature here is that a change in the distribution of the elastic
energy takes place, from nearly uniform before the transformation to a
pattern where the energy is concentrated on the ensuing ETD, and where
the bulk portion of the cell relaxes to a configuration of much
smaller elastic strain. Relaxation of internal atomic coordinates,
after the transformation, leads to further reduction of the strain in
the cell. Finally, in Fig.~\ref{fig2} (c) the imposed homogeneous
strain is lifted and the system relaxes to a stress-free state, where
the elastic energy in the bulk portions of the cell nearly vanishes,
and most of the residual strain is located on the defect.

\subsection{Deformations of graphene containing small holes and 
topological defects}
\begin{figure}[b]
\includegraphics[width=8.5 cm]{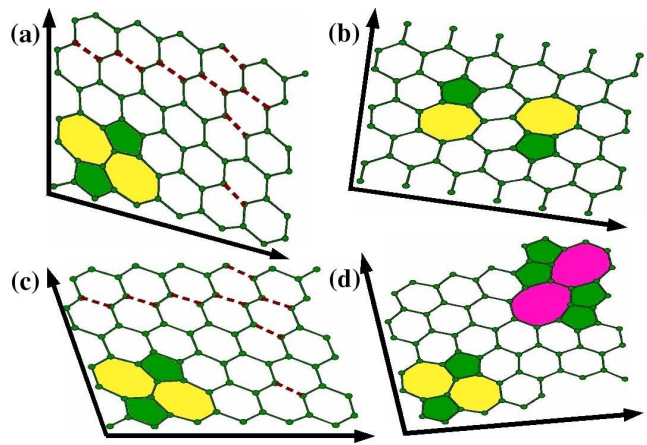}\\
\caption{Combination of homogeneous and inhomogeneous deformations of a
graphene sheet: (a) and (c) show two images of
homogeneously sheared supercells with a Stone Wales defect and
inhomogeneous deformations generating lines of strained bonds. (b)
Relaxation of the deformations in (a) leads to dislocation-dipole
(Stone Wales defect) dissociation. (d) Relaxation of the deformations
in (c) leads to formation of an extended topological defect.}
\label{fig3}
\end{figure}

We consider now more complex deformations of a graphene sheet,
including small sized holes generated by partially ripping a graphene
sheet, coupled with strongly inhomogeneous bond deformations. These
can be considered a model for severely strained and ripped graphene
samples, such as those obtained by reduction of GO.~\cite{cristina} In
the following, we show that healing of such graphene fragments,
combined with relaxation of the inhomogeneous deformations, also lead
to the formation of ETDs, in some cases with morphologies similar to
those that have been observed in recent experiments.

\subsubsection{Inhomogeneous deformations of graphene containing Stone-Wales defects}
We start with the supercell shown in Fig.~\ref{fig1}(b). By imposing a
homogeneous shear strain, coupled with lines of strained C-C bonds
(elongated by 27$\%$ to $\sim$1.8~\AA) as indicated by dashed segments
in Fig.~\ref{fig3}(a) and (c), and allowing for full relaxation of
both internal coordinates and homogeneous strain, the system reaches a
stress-free state where the dislocation dipole is dissociated into two
individual dislocations separated by a lattice constant, as shown in
Fig.~\ref{fig3}(b). When the strained bonds combine with a homogeneous
shear oriented with respect to the SW defect as shown in
Fig.\ref{fig3}(c), the SW defect induces the nucleation of an ETD
consisting of two side sharing octagons connected to two pairs of
pentagons, each on one of two opposite sides of the octagon pair,
shown in Fig.~\ref{fig3}(d). This ETD is similar in extent to those
resulting from the 555-777 and 585 reconstructions of a divacancy in
graphene.~\cite{kotakoski}

\subsubsection{Inhomogeneous deformations of topological defect 
networks in graphene}
\begin{figure}[b]
\includegraphics[width=8.5 cm]{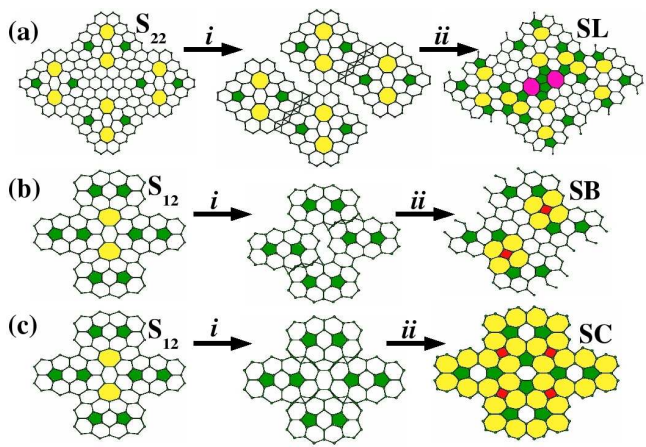}\\
\caption{Lowest-energy extended topological defects obtained from
regeneration of holes and response to inhomogeneous deformations.
Left column shows the parent geometry. Middle column shows imposed
deformation. Right column shows resulting stress-free metastable ETD
after full relaxation of internal atomic coordinates and cell vectors.
(a) Pristine (without functionalization) showing clusters of 5-7
pairs as well as a 8558 unit. (b) Hydrogen-functionalized and (c)
Hydroxyl-functionalized graphene show a morphological pattern where a
tetragon is surrounded by four heptagons, with pentagons connected to
its perimeter.}
\label{fig4}
\end{figure}

In order to further investigate the morphologies of deformed graphene,
and to access the role of functionalization and preexisting
topological defects, we choose as starting configurations the three
topological-defect graphene networks~\cite{jrc} shown on the left in
Fig.~\ref{fig4}, which were labelled $S_{12}$, $S_{31}$ and $S_{22}$
in Ref.~18. We also consider the graphane~\cite{elias}
(hydrogen-functionalized) and graphenol (hydroxyl-functionalized)
versions of these structures, both at full coverage, with graphenol
taken as a simple model for graphene oxide.~\cite{lerf} A variety of
deformation patterns, including small holes, as well as compressed
and/or elongated bonds were initially imposed on these structures, and
each system was allowed to relax (with optimization of both atomic
coordinates and supercell vectors) onto a stress-free metastable local
minimum of the total-energy surface.

Figure~\ref{fig4} shows starting configurations before and after
introduction of holes and inhomogeneous compressions, and the
resulting relaxed ETD configuration. In our calculations, these are,
respectively, the lowest-energy ETD for each of the following three
cases: (a) pristine graphene, (b) graphane, and (c) graphenol. The two
functionalized cases in Figs.~\ref{fig4}(b) and (c) show the formation
of a morphological pattern in the shape of a flower, where a square
ring is surrounded by four heptagons, with pentagons connected at the
perimeter. In the case of pristine graphene, the lowest-energy ETD we
obtained in our calculations contains pentagon-heptagon clusters
connected to pentagon-octagon units of the same morphology as
the one that has been recently observed as the periodic unit of a
domain-boundary in a graphene layer deposited on a nickel
substrate.~\cite{umesh,lahiri,ssa}

Further morphologies are obtained by considering the various
deformation patterns described above. While the formation of the ETDs
observed experimentally is largely dictated by kinetics, the
thermodynamic stability of such structures can be analyzed by
computing their formation energies, $E_f$. In the following, we
describe a total of 12 ETD configurations with $E_f\le
0.30$~eV/atom. To put these formation energies in perspective, we
quote the formation energy per atom of a fullerene molecule, 0.40
eV/atom with respect to a graphene sheet, and the energy of the
pentaheptite, a theoretical allotrope of graphene composed entirely of
pentagons and heptagons, 0.24 eV/atom,~\cite{crespi,jrc} both
computed with the same methodology as the ETDs energies in this work.

\subsubsection{Formation energy and energetics}
In the present study, the formation energies per atom of the
structures without functionalization are defined as
\begin{equation}
E_f = \frac{E_{ETD}}{N} - \mu_C\;;
\end{equation}
where $E_{ETD}$ is the {\it ab initio} total energy of the $N$-atom
supercell containing the ETD and $\mu_C = E_{graphene}/N$ is the total
energy per atom of an $N$-atom bulk graphene cell. For systems with
hydrogen functionalization, we take graphane as reference, such that
the formation energies per CH unit are defined as
\begin{equation}
E_f = \frac{E_{ETD+H}}{N} - (\mu_C + \mu_H)\;;
\end{equation}
where we keep $\mu_C$ as the energy per atom in graphene, and define
$\mu_H = \mu_{CH}-\mu_C$, where $\mu_{CH} = E_{graphane}/N$ is the
total energy per CH molecule for a bulk graphane calculation. For
the hydroxyl-functionalized cases, we have:
\begin{equation}
E_f = \frac{E_{ETD+OH}}{N} - (\mu_C + \mu_{OH})\;,
\end{equation}
for the energy per COH unit, where again we keep $\mu_C$ as the
energy per atom in graphene, and define $\mu_{OH} = \mu_{COH}-\mu_C$,
where $\mu_{COH} = E_{graphenol}/N$ is the total energy per COH
radical in a bulk graphenol calculation.
\begin{figure} [b]
\includegraphics[width=8cm]{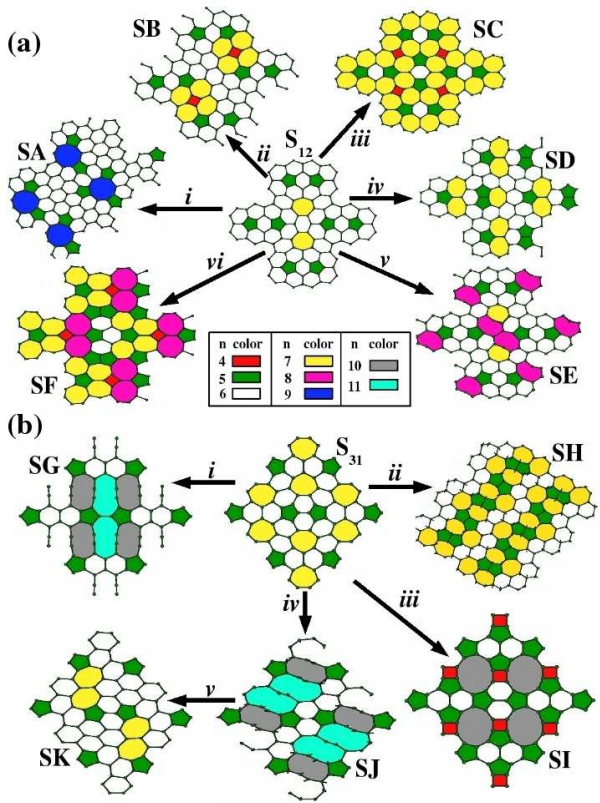}\\
\caption{(a) Low-energy extended topological defects derived from
deformed versions of the S$_{12}$ structure (shown in the
center). (b)Low-energy extended topological defects derived from
deformed versions of the S$_{31}$ structure (shown in the center at
the top row). Color coding for topological defects, ranging from
tetragons to hendecagons, is indicated.}
\label{fig5}
\end{figure}

Table I shows the formation energies of the 12 ETDs, along with the
energies of the parent geometries, $S_{12}$, $S_{31}$, and $S_{22}$,
both with and without functionalization.  In some cases, regeneration
of holes did not take place, a feature that is consistent with the
presence of holes in the final state of the RGO samples in
Ref.~5. These are indicated in Table I as
unreconstructed (UNREC.).
\begin{table}[t]
\caption{Formation energy $E_{f}$, in eV per formula unit, of extended
topological defects in graphene. $E_{g}$ is the electronic band gap in
eV. Structures without functionalization are metallic, except the
$S_{31}$. Functionalized structures are semiconductors or insulators,
except the SE.}  \centering
\begin{tabular} {|c|c|c|c|c|c|}
 \hline
& \multicolumn{3}{c}{$E_{f}$} \vline&  \multicolumn{2}{c}{$E_{g}$} \vline\\
\cline{2-6}
Structure &  & +H & +OH & +H & +OH  \\
\hline
graphene & 0.00 & 0.00 & 0.00  & 4.70 & 2.48 \\
\hline
$S_{12}$ & 0.40 & 0.27 & -  & 4.83 & - \\
\hline
$S_{22}$ & 0.26 & 0.16 & -  & 4.67 & - \\
\hline
$S_{31}$ & 0.36 & 0.29 & 0.36 &  5.15 & 0.51\\
\hline
SA & 0.27 & 0.22 & {\textit {UNREC.}}  & 4.97 & - \\
\hline
SB & 0.26 & 0.18 & 0.25 & 5.05 & 2.98\\
\hline
SC & 0.37 & 0.21 & 0.20 & 5.20 & 3.04 \\
\hline
SD & 0.32 & 0.26 & 0.35 & 5.14 & 2.94 \\
\hline
SE & 0.43 & 0.30 & 0.29 &\multicolumn{2}{c}{$\ast$} \vline \\
\hline
SF & 0.28 & 0.27 & {\textit {UNREC.}}& 5.29 & - \\
\hline
SG & 0.59 & 0.32 & 0.22 & 0.82 & 1.44 \\
\hline
SH & - & 0.24 &0.47 & 1.05 &2.59 \\
\hline
SI & 0.40 & 0.30 & 0.41 &1.30 & 2.75 \\
\hline
SJ & - & 0.34 & 0.28 &2.71 &0.43 \\
\hline
SK & 0.33 & 0.26 &0.27 & 1.80 & 0.84\\
\hline
SL & 0.24 & -  & - & - & - \\
\hline
\end{tabular}
\label{table-Energia}
\end {table}

Figure~\ref{fig5}(a) shows the six structures derived from deformed
versions of the S$_{12}$ geometry.~\cite{jrc} From the figure,
formation of pentagon-heptagons clusters is clearly the most common
morphological patterns emerging from small-hole regeneration and
relaxation of deformed graphene sheets, as found in recent
experiments, but other morphologies are also manifest. We observe
polygons ranging from tetragons to hendecagons, with color-coding
indicated in the figure. The ETD in structure SA, with a nonagon
connected to pentagons, is very similar to that present in the model
of amorphous graphene recently proposed as a high-specific area
material for supercapacitor applications.~\cite{yanwu}  The five
structures resulting from deformations of the S$_{31}$ geometry are
shown in Fig.~\ref{fig5}(b). Linear polyacetylene-like chains are
observed in the SG and SJ hydrogenated clusters, in structures with a
strong $sp^3$-hibridization content, which are stabilized by the
functional groups. Such polymer-like carbon chains were also obtained
in molecular dynamics models of RGO. Removal of functionalization in
the SJ geometry leads to the recovery of a threefold connected network
of pentagon-heptagon clusters, in the form of a climbed dislocation
dipole, of the type observed in irradiated graphene
samples.~\cite{meyer} Another striking geometry is the SI, which shows
the presence of tetragons connected to decagons, with a reasonably low
formation energy of 0.3 eV/atom when functionalized with
hydrogens. Note the presence of a network of large pores, which would
make defective graphene sheets with such morphologies candidates for
applications as supercapacitors~\cite{yanwu} and selective permeable
membranes.~\cite{schrier,enjiang}

\subsubsection{Role of functionalization}
Regarding the role of functionalization, we observe in Table I that
the formation energy decreases with functionalization in all cases,
with the exception of the hydroxyl-functionalized S$_{31}$, SD, and
SI, which have slightly higher values of $E_f$ than their pristine
counterparts. For the hydrogen-functionalized ETDs, an average
reduction of 95~meV/formula-unit is observed in Table I, while for the
hydroxyl-functionalized ETDs, we obtain an average reduction of
125~meV/formula-unit. In both cases rather sizable maximum reductions
of $E_f$ of 0.27 and 0.37 eV/formula-unit occurs, respectively, for
the H- and OH-functionalized SG morphologies, where functionalization
leads to stabilization of polymer-like carbon chains.
Functionalization also changes the electronic structure: while all
pristine ETDs are metallic, with the exception of the S$_{31}$ parent
geometry, the functionalized ETDs are all semiconductors or
insulators, with the exception of the SE.  Table I shows the values of
the gap for the semiconducting and insulating ETDs.  This change in
the electronic structure is connected with the incorporation of an
$sp^3$ component in the electronic structure of these graphene
materials, associated to the bonding of the functional groups to the
graphene sheet.

\section{Conclusion}
In conclusion, ab initio calculations indicate that stress-relaxation
of a graphene sheet containing topological defects may lead to the
formation of extended topological defects showing morphological units
that have been observed in recent experiments. We uncover the role of
topological defects as a stress-accumulation site that induces the
bond rotation events that generate the ETDs. Furthermore, we find that
healing of small voids coupled with the relaxation of inhomogeneously
strained regions also lead to the formation of ETDs, revealing a rich
variety of morphological patterns and plastic deformation mechanisms
in pure and functionalized graphene structures. Our results indicate
that the tendency of deformed graphene sheets to form topological
defect clusters is enhanced in the presence of functional groups, with
a systematic reduction of the formation energies of such defective
structures in graphane and in graphenol, when compared to graphene, in
agreement with the available experimental evidence.

\begin{acknowledgments}
The authors acknowledge support from the Brazilian agencies CNPq,
FAPEMIG, Rede de Pesquisa em Nanotubos de Carbono, INCT de
Nanomateriais de Carbono, and Instituto do Mil\^enio em
Nanotecnologia-MCT.
\end{acknowledgments}

\end{document}